\documentstyle[12pt]{article} 

\newcommand{\bea}{\begin{eqnarray}}
\newcommand{\eea}{\end{eqnarray}}

\begin{document}

\begin{titlepage}

\begin{flushright}
ITP-SB-97-72 \\
hep-th/9712027 \\
\end{flushright}

\begin{center}
\vskip3em
{\large\bf Comment on the Bound State Problem in $N=4$ Super Yang-Mills Theory}

\vskip3em
{Gordon Chalmers\footnote{E-mail:chalmers@insti.physics.sunysb.edu}\\
\vskip .5em
{\it Institute for Theoretical Physics\\ State University of New York\\
 Stony Brook, NY 11794-3840, USA\\}
\vskip2em
}
\end{center}

\vfill

\begin{abstract}  

We re-examine the threshold bound state problem on the wrong sign 
Taub-Nut space; the metric on which describes the relative moduli space 
of well separated BPS monopoles.  The quantum mechanics   
gives rise to a continuous family of threshold bound states, in distinction 
to the unique one found on the Atiyah-Hitchin metric.  
\end{abstract}

\vfill
\end{titlepage}


\section{Introduction}\label{intro}

The quantum aspects of monopole dynamics has been under much study
in the past few years to a large extent because of the evidence
found for $S$-duality in $N=4$ supersymmetric Yang-Mill theory.
In particular, Sen first produced evidence \cite{sen} for
the existence of a threshold bound state of magnetic charge two
in a spontaneously broken $SU(2)$ $N=4$ supersymmetric Yang-Mills
theory.  Such states
are found by looking for smooth normalizable ground states of the
quantum mechanics model describing the slowly moving magnetic
monopoles.

In this note we re-examine this problem in the limit where we
decouple the massive vector boson while maintaining the monopole
degree of freedom (i.e. $m_W=vg\rightarrow\infty$ and $m_m={v/g}
\rightarrow {\rm constant}$).  We point out that the quantum
mechanics of point-like dyons is smooth, even in the case where
the equations of motion governing the particles in this
approximation are singular.  We show that the Hamiltonian which arises
within the low-energy approximation to the dynamics of
point-like dyons possesses a one-parameter family of smooth ground
states.  This enlargement in the number of threshold bound
states is surprising because the moduli space metric is singular.

\vskip .4in
\section{Qauntum Mechanics of Point-like dyons}

The slow-motion dynamics of monopoles is governed by geodesic
motion derived from the metrics on particular hyper-K\"ahler manifolds 
\cite{AH}; although in principle the
metrics on these spaces are difficult to find explicitly, approximate
forms describing the asymptotic regime may be found through a
Lagrangian description \cite{M}.  Namely, in the limit of long-range
interactions, where
the effects of the $W$-bosons are neglected we may use the well known 
world-line
formulation of point-like dyons to derive the forces (and
corresponding metric) \cite{wy1,wy2}.  In the
following we reanalyze the classic work of Sen \cite{sen} in
this approximation.

The two-monopole moduli space metric is uniquely determined to be 
the Atiyah-Hitchin metric \cite{AH}.  Its asymptotic form limits to the double 
cover of the Taub-NUT metric, which possesses a tri-holomorphic isometry reflecting 
the fact that there is an additional charge conservation.  The equations 
of motion describing the time evolution of the zero modes is derived from the 
bosonic quantum mechanics problem,  
\bea 
{\cal L}= \int dt~ g^{ab} {\dot x}_a {\dot x}_b  \ .
\label{mapp}
\eea
The Taub-NUT metric is explicitly :
\bea
ds^2=(1-{1\over r}) d{\vec r}\cdot d{\vec r} + {r \over r-1}
 (d\psi + \vec{\omega}\cdot d{\vec r})^2 \ ,
\label{TN}
\eea
which has a real singularity at $r=1/m_w$, the Compton wavelength
of the $w$-boson (in the remainder of this note we absorb the mass parameter 
into the radial coordinate).  Including the $w$-bosons, in which
non-abelian monopoles are described, smooths out the singularity appearing in
the charge-rotator degree of freedom; in this case one
obtains the Atiyah-Hitchin metric \cite{AH}.

The quantum mechanics describing the system (\ref{mapp}) possesses
states in one-to-one correspondence with the Hamiltonian found
from the Lagrangian (\ref{mapp}) \cite{w}.  We first introduce the vielbein 
basis
\bea
v^0=\bigl(1-{1\over r}\bigr)^{1/2} dr \qquad
v^1=r\bigl(1-{1\over r}\bigr)^{1/2} \sigma_1
\eea
\bea
v^2=r\bigl(1-{1\over r}\bigr)^{1/2} \sigma_2 \qquad
v^3=-\bigl(1-{1\over r}\bigr)^{-1/2} \sigma_3  \ ,
\eea
in which the metric may be expressed as $g=\sum_{j=0}^3 v^j\otimes v^j$.
The Hamiltonian written in terms of these is simply 
\bea
H=d^\dagger d+ d d^\dagger \ ,
\label{ham}
\eea 
and the quantum states are in correspondence with the eigenfunctions of
the Hamiltonian.  In the following we examine only the $2$-form eigenfunctions 
of the Hamiltonian.  

In terms of $v_i$, we make an ansatz for the self-dual $2$-forms which are 
annihilated by $H$ with 
\bea
\omega_i^{\pm} = F_i (r)\Bigl( v^0\wedge v^i \pm
  {1\over 2} \epsilon^i_{jk} v^j\wedge v^k \Bigr) \ .
\eea  
We first introduce the left-invariant $1$-forms of $SO(3)$, $\sigma_j$.   
They satisfy $d\sigma_i={1\over 2} \epsilon_{ijk} \sigma^j \wedge \sigma^k$ 
and have an explicit form:
\bea 
\sigma_1 = -\sin\psi d\theta + \cos\psi \sin\theta d\phi 
\eea 
\bea 
\sigma_2 = \cos\psi d\theta + \sin\psi \sin\theta d\phi 
\eea 
\bea
\sigma_3 = d\psi + \cos\theta d\phi \ .
\eea  
On wrong-sign Taub-NUT we find the $2$-form solutions to the Hamiltonian 
in (\ref{ham}), 
\bea
\omega_1^{+} = e^{-r}\Bigl\{ (r-1)dr\wedge\sigma_1
 - r d\sigma_1\Bigr\}
\eea
\bea
\omega_1^{-} = {1\over r^2} e^{r}\Bigl\{ (r-1)dr\wedge\sigma_1
 + r d\sigma_1\Bigr\}
\eea
\bea
\omega_2^{-} = {r^2\over (r-1)^4}
  \Bigl\{ dr\wedge\sigma_2+r^2(1-{1\over r}) d\sigma_2 \Bigr\}
\eea
and
\bea
\omega_2^{+} = {1\over (r-1)^2}
  \Bigl\{ -dr\wedge\sigma_2+r^2(1-{1\over r}) d\sigma_2 \Bigr\} \ .
\eea  
The solutions to $\omega_3^{\pm}$ follow by replacing $\sigma_2$ with 
$\sigma_3$.  

The solutions for $\omega_2^{\pm}$ are found by replacing $\sigma_1$
with $\sigma_2$ in $\omega_1^{\pm}$.  All of the two-forms obey
\bea
\omega_i^{(\alpha)}\wedge\omega_j^{(\beta)}=0  \ ,
\eea
for $(i,\alpha)\neq (j,\beta)$.  Only the solutions $\omega_1^+$
and $\omega_2^+$ are smooth and normalizable.  They both may
serve as zero-energy ground states of the Hamiltonian (\ref{ham}).

Both $\omega_1^+$ and
$\omega_2^+$ are odd under $\psi\rightarrow \psi+\pi$.  They
would correspond to odd electrically charged states if interpreted
as threshold bound states of the abelian monopole system.  Further,
under $\psi\rightarrow\psi+\theta$ they change as :

\bea
\Bigl(\matrix{\omega_1^+ \cr \omega_2^+} \Bigr)\rightarrow
  \Bigl(\matrix{\cos\theta & \sin\theta \cr -\sin\theta & \cos\theta} \Bigr)
  \Bigl(\matrix{\omega_1^+ \cr \omega_2^+ }\Bigr)
\eea

The Hamiltonian (\ref{ham}) possesses a family of normalizable ground
states of the form
\bea
\omega_\alpha =\cos{\theta} ~\omega_1^+ + \sin{\theta}~ \omega_2^+ \ ,
\eea
where both $\omega_i^\pm$ are normalized to one.  In a complex basis
\bea
\omega=\omega_1^+ + i \omega_2^+ \qquad
  {\bar\omega} = \omega_1^+ - i \omega_2^+  \ ,
\eea
these states are written as
\bea
\omega_\alpha = e^{i\alpha} \omega + e^{-i\alpha} {\bar\omega}   
\label{relbound} \ .
\eea 
The harmonic $2$-forms $\omega^\pm$ both are odd under shifts of the coordinate 
$\theta$ by $\pi$.  The total moduli space for the two monopoles has the 
form $M_2= R^3 \otimes {S^1\otimes M_2\over Z_2}$, where we have approximated to only the 
asymptotic region (i.e. the Taub-NUT metric).  The complete wave function is found
by tensoring the relative bound states $\omega^\pm$ with the wave function 
on center of mass component $R^3$
of the total moduli space.  We see, according to the 
analysis of Sen \cite{sen}, that the states in (\ref{relbound}) correspond to dyonic 
bound states with electric charge $\pm 1$: the negative (positive) electrically 
charged bound states are $\omega^+$ ($\omega^-$) as may be seen under a continuous 
rotation of the $\theta$ coordinate.   

The existence of these bound states is surprising for two reasons.  First, the 
zero energy eigenstates are smooth and normalizable which means that despite the 
singularity on the wrong sign Taub-NUT the quantum mechanics is sensible.  Second, 
the appearance of a $U(1)$ set of bound states appears to be in contradiction with 
the $SL(2,Z)$ invariance of the $N=4$ supersymmetric Yang-Mills spectrum.  

The spectrum of $N=4$ super Yang-Mills theory is invariant under $SL(2,Z)$ 
transformations, which takes a BPS soliton with magnetic and electric 
quantum numbers $(m,n)$ to 
\bea   
\Bigl(\matrix{m \cr n} \Bigr)\rightarrow
  \Bigl(\matrix{p & q \cr r & s} \Bigr)
  \Bigl(\matrix{m \cr n}\Bigr)
\eea 
with $pr-sq=1$.  In the limit we are taking all of the BPS states satisfying the mass 
formula, $m^2=2g\vert n_m v_D + n_e v\vert^2$ ($v_D=v/g$), with non-vanishing 
electric quanta decouple.  Under the $S$-type duality transformation the coupling 
constant gets mapped from $g\rightarrow 1/g$ and the decoupled electric states 
change to purely magnetic ones with zero mass.  Our $U(1)$ set of bound states 
reflects a condensation of an infinite tower of massless magnetic states.

However, in $N=4$ super Yang-Mills theory 
there are two distinct phases, namely $v=0$ and $v\neq 0$.  In the analysis of the 
threshold bound state problem originally done in \cite{sen}, one works in the latter 
case:  In the Atiyah-Hitchin metric there is only one mass scale and 
the (non-zero) vacuum value of the scalar field may be eliminated via a coordinate 
transformation.  The threshold bound state calculated in \cite{sen} and its 
implications towards the $SL(2,Z)$ duality symmetry of $N=4$ super Yang-Mills 
theory is valid only in this case, as opposed to the pure abelian system considered 
here.

\section{Discussion}  

We have re-examined the threshold bound state problem first analyzed by Sen \cite{sen}, 
but for the abelian monopole system.  In the two-monopole sector there arises a 
continuous family of bound states with magnetic charge two and electric charge $\pm 1$; 
such states do not violate the $SL(2,Z)$ duality symmetry which exists in a 
completely broken $N=4$ super Yang-Mills theory at finite coupling.  Similar 
bound states exist on the higher dimensional analogs of the Taub-Nut metric.  In 
future work we plan to investigate the relation of the enlargement of the 
number of threshold bound states to the problem of enhanced gauge symmetry of 
the Yang-Mills system describing coincident $p$-branes \cite{w2}.

\vskip 1em
\noindent{\large\bf Acknowledgements}
\vskip 1em

\noindent{Gordon Chalmers was supported by NSF grant No.~PHY 9309888.}

\end{document}